\documentclass{aa} 

\usepackage{graphicx}
\usepackage{longtable}
\usepackage{txfonts}
\usepackage{color}
\usepackage{amstext}
\begin{document}

\title{Using non-solar-scaled opacities to derive stellar parameters}
   
\subtitle{Toward high-precision parameters and abundances}
\titlerunning{Non-solar-scaled stellar parameters}
\authorrunning{Saffe et al.}

\author{C. Saffe\inst{1,2,5}, M. Flores\inst{1,2,5}, P. Miquelarena\inst{2}, F. M. L\'opez\inst{1,2,5},
M. Jaque Arancibia\inst{1,4,5}, A. Collado\inst{1,2,5}, E. Jofr\'e\inst{3,5} \and R. Petrucci\inst{3,5}}

\institute{Instituto de Ciencias Astron\'omicas, de la Tierra y del Espacio (ICATE-CONICET), C.C 467, 5400, San Juan, Argentina.
               \email{[csaffe,matiasflorestrivigno,fmlopez,mjaque]@conicet.gov.ar}
         \and Universidad Nacional de San Juan (UNSJ), Facultad de Ciencias Exactas, F\'isicas y Naturales (FCEFN), San Juan, Argentina.
         \and Observatorio Astron\'omico de C\'ordoba (OAC), Laprida 854, X5000BGR, C\'ordoba, Argentina.
               \email{[emiliano,romina]@oac.unc.edu.ar}
         \and Departamento de F\'isica y Astronom\'ia, Universidad de La Serena, Av. Cisternas
         1200. 1720236, La Serena, Chile.
        \and Consejo Nacional de Investigaciones Cient\'ificas y T\'ecnicas (CONICET), Argentina
         }

\date{July 2018}

 
\abstract
{}
{ In an effort trying to improve spectroscopic methods of stellar parameters determination,
we implemented non-solar-scaled opacities in a simultaneous derivation of fundamental
parameters and abundances. We want to compare the results with the usual solar-scaled method using a sample
of solar-type and evolved stars. }
{We carried out a high-precision stellar parameters and abundance determination by applying 
non-solar-scaled opacities and model atmospheres. Our sample is composed by 20 stars (including
main-sequence and evolved objects), with six stars belonging to binary systems.
The stellar parameters were determined by imposing ionization and excitation
equilibrium of Fe lines, with an updated version of the FUNDPAR program, together with plane-parallel ATLAS12
model atmospheres and the MOOG code. Opacities for an arbitrary composition and v$_{\rm{micro}}$ were calculated
through the OS (Opacity Sampling) method.
Detailed abundances were derived using equivalent widths and spectral synthesis with the MOOG program.
We applied the full line-by-line differential technique using the Sun as reference star, both in the
derivation of stellar parameters and in the abundance determination.
We start using solar-scaled models in a first step, and then continue the process but scaling to the
abundance values found in the previous step (i.e. non-solar-scaled).
The process finish when the stellar parameters of one step are the same of the previous step, i.e.
we use a doubly-iterated method.}
{We obtained a small difference in stellar parameters derived with non-solar-scaled
opacities compared to classical solar-scaled models.
The differences in T$_{\rm{eff}}$, {log g} and [Fe/H] amount up to 26 K, 0.05 dex and 0.020 dex
for the stars in our sample.
These differences could be considered as the first estimation of the error due to the use of 
classical solar-scaled opacities to derive stellar parameters with solar-type and evolved stars.
We note that some chemical species could also show an individual variation higher than those of the [Fe/H] 
(up to $\sim$0.03 dex) and varying from one specie to another, obtaining a chemical pattern difference
between both methods. This means that condensation temperature T$_{c}$ trends could also present a variation.
We include an example showing that using non-solar-scaled opacities, the solution found with the
classical solar-scaled method indeed cannot always verify the excitation and ionization balance conditions
required for a model atmosphere.
We discuss in the text the significance of the differences obtained when using solar-scaled vs non-solar-scaled
methods.
}
{
We consider that the use of the non-solar-scaled opacities is not mandatory e.g. in every statistical
study with large samples of stars. However, for those high-precision works whose results depend on the mutual
comparison of different chemical species (such as the analysis of condensation temperature T$_{c}$ trends),
we consider that it is whortwhile its aplication.
To date, this is probably one of the more precise spectroscopic methods of stellar parameters derivation.
}


   \maketitle
%

\section{Introduction}

The discovery of the first exoplanets orbiting the pulsar PSR1257+12 \citep{wolszczan-frail92}
and the solar-type star 51 Peg \citep{mayor-queloz95}, gave rise to a number of works which
strongly motivated to increase the precision of both photometry and spectroscopy techniques.
This continuous effort allowed the discovery of new planets and its subsequent analysis.
For instance, radial-velocity measurements improved to a precision of few m/s or less 
\citep[e.g.][]{locurto15,fischer16}, while the {\it{Kepler}} photometry could reach < 1 millimag 
for a 12$^{th}$ mag star\footnote{https://keplergo.arc.nasa.gov/pages/photometric-performance.html}.
The derivation of detailed chemical abundances followed a similar path.
For example, the use of the called differential technique applied to physically similar stars,
allowed to significantly reduce the dispersion in [Fe/H] to values near or lower than $\sim$0.01 dex
\citep[e.g.][]{desidera04,melendez09,ramirez11,saffe15,saffe16,saffe17}. These high precision values are needed,
for example, to detect a possible chemical signature of planet formation \citep[e.g.][]{melendez09,saffe16}
and also required by the chemical tagging technique.
Then, it is crucial to pursuit the maximum possible precision in the derivation of stellar parameters and
chemical patterns.

Several works studying the chemical composition of solar-type stars use a two-step method of
abundance determination.
For instance, in the study of stellar galactic populations \citep[e.g. ][]{adibekyan12,adibekyan13,adibekyan14,adibekyan16,mena17},
metallicity trends in stars with and without planets \citep[e.g.][]{sousa08,sousa11a,sousa11b,adibekyan12},
and the possible signature of terrestrial planets \citep[e.g.][]{gonz-hern10,gonz-hern13,adibekyan14}.
These works are not an exhaustive list but exemplify a number of important studies and trends.
Briefly, in a first step the fundamental parameters are determined by imposing excitation
and ionization equilibrium of Fe I and Fe II lines.
The model atmosphere which satisfy these conditions is usually interpolated or calculated
by assuming both solar-scaled opacities and abundances.
Once fixed the stellar parameters, in a second step the chemical abundances are determined
by using equivalent widths or spectral synthesis, depending on the possible presence
of blends and other effects such as hyperfine structure (HFS).
The process normally finishes here, resulting in a chemical pattern that 
is not exactly solar-scaled, as supposed in the first step.
Notably, even reaching a perfect match between synthetic and observed spectra, this inconsistency
could lead to an incorrect determination of stellar parameters and abundances.
In addition, this issue is generally unaccounted in the total error estimation of most literature works.

Then, in an effort to improve the precision of the results, the approppriate calculation of the model
atmospheres should include the previous derivation of opacities obtained for a specific
abundance pattern, beyond the classical solar-scaled values.
We wonder if it is possible to implement such method in the calculation of stellar parameters
in a practical way.
What is the difference in the stellar parameters obtained with this procedure and the classical solar-scaled
methods? How many iteration steps are neccessary to properly derive
the stellar parameters? Can this effect introduce a statistical bias in studies of large samples?
Do we expect a null difference in metallicity between very similar components of binary stars?
These important questions are the motivation of the present work.

This work is organized as follows. In Sect. 2 we describe the sample and data reduction,
while in Sect. 3 we explain the calculation of opacities and models. Finally, we present 
the discussion and conclusions in Sect. 4 and 5.

\section{Sample of spectra}

As previously mentioned, the detection of the possible chemical signature of planet
formation requires a very high precision in stellar parameters.
These studies are usually performed on solar-type main-sequence stars
\citep[e.g.][]{tucci-maia14,saffe15,saffe16}.
We start by choosing 10 objects of this type for our sample.
In order to study the possible differences between solar-scaled and non-solar-scaled
methods in other type of stars, we also included in our sample 10 giants stars (see Table \ref{tab.samples}).
Then, the final sample is composed by 20 stars with T$_{\rm{eff}}$ in the range {4131-8333} K and
log g in the range {1.62-4.64} dex.
Their metallicities range from $-$0.43 to $+$0.27 dex i.e. including objects with values lower
and higher than the Sun, and likely including a number of different chemical mixtures.
Six stars in our sample belong to binary systems
previously studied in the literature \citep[][hereafter SA15, SA16, SA17]{saffe15,saffe16,saffe17}.
In particular, the binary systems were selected due to a high degree of physical similarity
between their components, which allow to test the possible variation
of the stellar parameters (scaled vs non-solar-scaled methods) depending on the chemical pattern,
which is roughly similar for the two stars in each binary system.

For the stars studied in this work, Table \ref{tab.samples} presents the corresponding spectrograph,
resolving power, signal to noise and object used as a proxy for the Sun spectra.
We reduced the data by using the reduction package MAKEE 3 with HIRES spectra\footnote{http://www.astro.caltech.edu/~tb/makee/},
the DRS pipeline (Data Reduction Software) with HARPS
data\footnote{https://www.eso.org/sci/facilities/lasilla/instruments/harps/doc.html} 
and the OPERA 5 \citep{martioli12} software with GRACES spectra.
The continuum normalization and other operations (such as Doppler correction and combining spectra)
were perfomed using Image Reduction and Analysis Facility (IRAF)\footnote{IRAF is distributed by the National Optical Astronomical
Observatories, which is operated by the Association of Universities for
Research in Astronomy, Inc. under a cooperative agreement with the
National Science Foundation.}.

\begin{table}
\centering
\caption{Sample of spectra used in this work.}
\scriptsize
\begin{tabular}{lccrc}
\hline
\hline
Stars & Spectrograph & R & S/N & Sun \\
      &              &   &     &spectra\\
\hline
Main-sequence stars \\
\hline
HD 80606 + HD 80607                & HIRES  & 67000  & $\sim$330 & Iris \\
$\zeta^{1}$ Ret + $\zeta^{2}$ Ret  & HARPS  & 110000 & $\sim$300 & Ganymede \\
HAT-P-4 + TYC 2567-744-1           & GRACES & 67500  & $\sim$400 & Moon \\
HD 19994                           & HARPS  & 115000 & $\sim$390 & Ganymede \\
HD 221287                          & HARPS  & 115000 & $\sim$130 & Ganymede \\
HD 96568                           & HARPS  & 115000 & $\sim$350 & Ganymede \\
HD 128898                          & HARPS  & 115000 & $\sim$340 & Ganymede \\
\hline
Evolved stars\\
\hline
HD 2114                            & HARPS  & 115000 & $\sim$180 & Ganymede \\
HD 10761                           & HARPS  & 115000 & $\sim$195 & Ganymede \\
HD 28305                           & HARPS  & 115000 & $\sim$180 & Ganymede \\
HD 32887                           & HARPS  & 115000 & $\sim$245 & Ganymede \\
HD 43023                           & HARPS  & 115000 & $\sim$275 & Ganymede \\
HD 50778                           & HARPS  & 115000 & $\sim$140 & Ganymede \\
HD 85444                           & HARPS  & 115000 & $\sim$225 & Ganymede \\
HD 109379                          & HARPS  & 115000 & $\sim$335 & Ganymede \\
HD 115659                          & HARPS  & 115000 & $\sim$310 & Ganymede \\
HD 152334                          & HARPS  & 115000 & $\sim$195 & Ganymede \\
\hline
\end{tabular}
\normalsize
\label{tab.samples}
\end{table}

\section{Calculating non-solar-scaled opacities and models}

Within the suite of Kurucz's programs for model atmosphere calculation,
a non-solar-scaled model can be calculated mainly in two different ways.
The first option consists in the previous calculation of an Opacity Distribution
Function (ODF) which depends on the abundances and microturbulence velocity v$_{\rm{micro}}$,
which is then used as input for an ATLAS9 model atmosphere \citep[see e.g.][]{kurucz74,castelli05b}.
The second option consists directly in the calculation of an ATLAS12 model atmosphere,
which calculates internally the opacities through the Opacity Sampling (OS) method
\citep[see e.g.][]{peytremann74,kurucz92,castelli05a}.

We will briefly describe both options here.
ODF functions used by ATLAS9 are tables which describe the dependence of the line absorption coefficient l$_{\nu}$
as a function of the frequency $\nu$, calculated for a given pair T$_{gas}$ and P$_{gas}$ \citep[see e.g.][]{kurucz74,castelli05b}.
Different sets of ODFs are derived for fixed abundances (usually solar-scaled and some
$\alpha$-enhanced models) and fixed v$_{\rm{micro}}$ \citep[see e.g.][]{castelli-kurucz03,coelho14}.
Strictly speaking, the calculation of an ATLAS9 model atmosphere for an specific composition and v$_{\rm{micro}}$
should include the previous calculation of the ODF for the corresponding values using the DFSYNTHE program
\citep[see e.g.][]{castelli05b}.
Otherwise, the numerical abundances derived for the ions during the actual model calculation at
a given gas state (T$_{gas}$ and P$_{gas}$) will differ from the ones calculated during the ODF computation.
On the other hand, it is also possible to derive an ATLAS12 model atmosphere in which the OS method determines
the opacity for given abundances and v$_{\rm{micro}}$. 
This is an important detail, given that the correct derivation of a model atmosphere for an
arbitrary chemical pattern should include the specific opacities, and not only a mere change
in the abundances used. The longer time required by the calculation of an ATLAS12 model
compared to ATLAS9 is not a problem today, finishing the operation after only few minutes.
There is available a complete ATLAS12 version for {\it{gfortran}} at the
Fiorella Castelli's webpage\footnote{http://wwwuser.oats.inaf.it/castelli/}, while
for DFSYNTHE there is only an ifort (intel) version. Then, we use ATLAS12 for the calculation
of a non-solar-scaled model atmosphere.

Usually, ATLAS12 model atmospheres are used when the chemical composition
of the stars present e.g. alpha-elements patterns that do not follow solar-scaled
or alpha-enhanced models, or early-type stars where diffussion effects change
significatively their superficial composition
\citep[see e.g.][]{sbordone05,mucciarelli12}.
In this work, we use non-solar-scaled opacities 
(calculated by ATLAS12) for the simultaneous derivation of both stellar parameters
and abundances, as we explain in the next Section. 

\section{Stellar parameters and chemical abundance analysis}

We started by measuring the equivalent widths of Fe I and Fe II lines in the spectra
using the IRAF task {\it{splot}}, and then continued with other
chemical species. The lines list and relevant laboratory data (such as excitation potential and
oscilator strengths) are similar to those used in previous works (SA15, SA16).
However we note that the exact values are not very relevant here, given the differential technique
applied in this case.

The first estimation of the stellar parameters (T$_{\rm{eff}}$, {log g}, [Fe/H], v$_{\rm{micro}}$) uses an
iterative process within the FUNDPAR program \citep{saffe11}, searching for a model atmosphere
which satisfies the excitation and ionization balance of Fe I and Fe II lines.
This code was improved in order to use the program MOOG \citep{sneden73} together with
ATLAS12 opacities and model atmospheres \citep{kurucz93}.
The procedure uses explicitly calculated (i.e. non-interpolated) plane-parallel local thermodynamic equilibrium
(LTE) Kurucz's model atmospheres with ATLAS12, which includes the internal calculation of the
line opacities through the Opacity Sampling (OS) method.
Two runs of ATLAS12 are used \citep[see e.g.][]{castelli05a}: the first one for a preselection of important
lines (for the given stellar parameters and abundances) and the second for the final calculation of the
model structure. In both runs we explicitly use a specific chemical pattern.

The models are calculated with overshooting in order to facilitate the comparison
with previous works \citep[e.g.][]{saffe15,saffe16,saffe17}. However, we caution that 
this modification of the mixing-length theory is suitable for the Sun but
not neccesarily for other stars \citep{castelli97}.
The overshooting produces a thermal structure that disagrees with that obtained from
hydrodynamical simulations, being magnified for low metallicity and dissapears
for solar metallicity \citep[see Fig. C.1 in ][]{bonifacio09}.
We adopted the layer 36 in the atmosphere
where numerical results related with the Schwarzschild criterion can be assumed as reliable
for models with T$_{\rm{eff}}$ $>$ 4000 K \citep{castelli05a}.

We applied in this work the full differential technique i.e. we consider
the individual line-by-line differences between each star and the Sun,
which is used as the reference star. The object used as a proxy for the solar
spectra (in reflected light) is listed in the last column of Table \ref{tab.samples}.
First, we determined absolute abundances for the Sun using 5777 K for T$_{\rm{eff}}$,
4.44 dex for {log g} and an initial v$_{\rm{micro}}$ of 1.0 km/s. Then, we estimated
v$_{\rm{micro}}$ for the Sun with the usual method of requiring zero slope in the absolute
abundances of Fe I lines vs. reduced equivalent widths and obtained a final v$_{\rm{micro}}$
of 0.91 km/s. We note, however, that the exact values are not crucial for our strictly
differential study \citep[see e.g.][]{bedell14,saffe15}.
The next step is the derivation of stellar parameters for all stars in our 
sample using the Sun as reference i.e. (star - Sun).
We want to stress that this line-by-line method is applied both in the
derivation of stellar parameters and in the chemical abundances
(i.e. the "full" differential technique), allowing to increase
the precision of stellar parameters \citep[e.g.][]{saffe15,saffe16}.

In the first FUNDPAR iteration we use solar-scaled models for an initial estimation of stellar parameters.
Then, a starting set of chemical abundances are determined using equivalent widths and spectral synthesis.
In the next step, the iterative process within FUNDPAR is restarted but scaling using the last
set of abundances instead of the initial solar-scaled values. At this point, we take advantage of the
ATLAS12 opacities and models calculated "on the fly" at the request of FUNDPAR, 
for an arbitrary chemical composition and v$_{\rm{micro}}$.
We note that the possibility to scale to an arbitrary chemical pattern (through an ABUNDANCE SCALE 
control card) was implemented by Dr. Fiorella Castelli and is not available in the original 
Kurucz code \citep{castelli05a}. In this way, new stellar parameters and abundances are succesively derived, finishing the
process consistently when the stellar parameters are the same of the previous step.
The process described here is then a doubly-iterated process, with the FUNDPAR iteration contained within
a larger (abundance-scaled) iteration. The classical solar-scaled results correspond to the values derived
in the first step of this new scheme.

We computed the individual abundances for the following elements:
\ion{C}{I}, \ion{O}{I}, \ion{Na}{I}, \ion{Mg}{I}, \ion{Al}{I}, \ion{Si}{I}, \ion{S}{I}, \ion{Ca}{I}, 
\ion{Sc}{I}, \ion{Sc}{II}, \ion{Ti}{I}, \ion{Ti}{II}, \ion{V}{I}, \ion{Cr}{I}, \ion{Cr}{II}, \ion{Mn}{I}, 
\ion{Fe}{I}, \ion{Fe}{II}, \ion{Co}{I}, \ion{Ni}{I}, \ion{Cu}{I}, \ion{Sr}{I}, \ion{Y}{II}, and \ion{Ba}{II}.
The HFS splitting was considered for \ion{V}{I}, \ion{Mn}{I}, \ion{Co}{I}, \ion{Cu}{I}
and \ion{Ba}{II}, by adopting the HFS constants of \citet{kurucz-bell95} and performing spectral synthesis
with the program MOOG \citep{sneden73} for these species.
We used exactly the same lines (laboratory data and equivalent
widths) for iron and for all chemical species, when deriving stellar parameters
and abundances with both methods (scaled and non-solar-scaled).

\section{Results and Discussion}

\begin{figure}
\centering
\includegraphics[width=8cm]{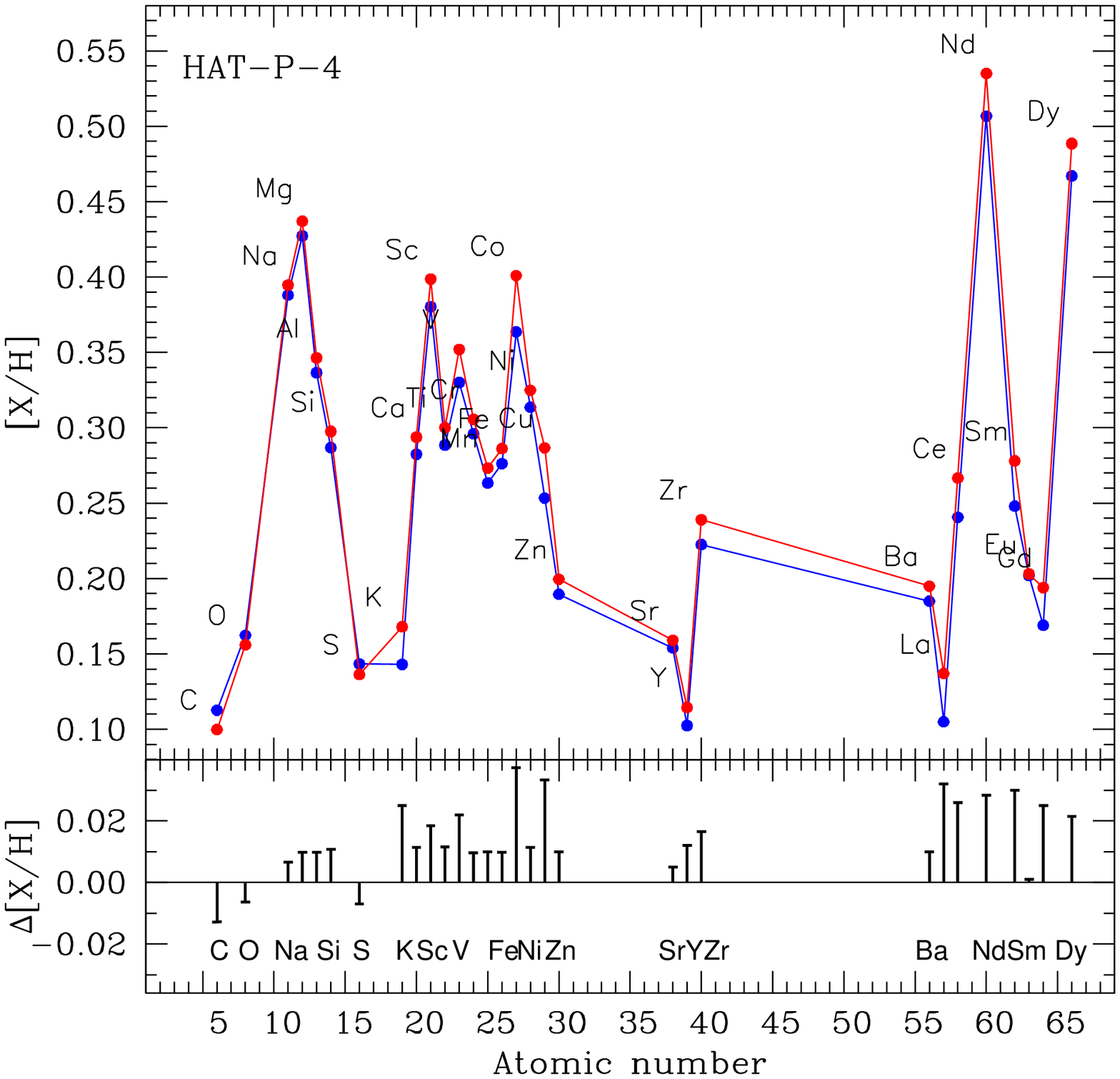}
\includegraphics[width=8cm]{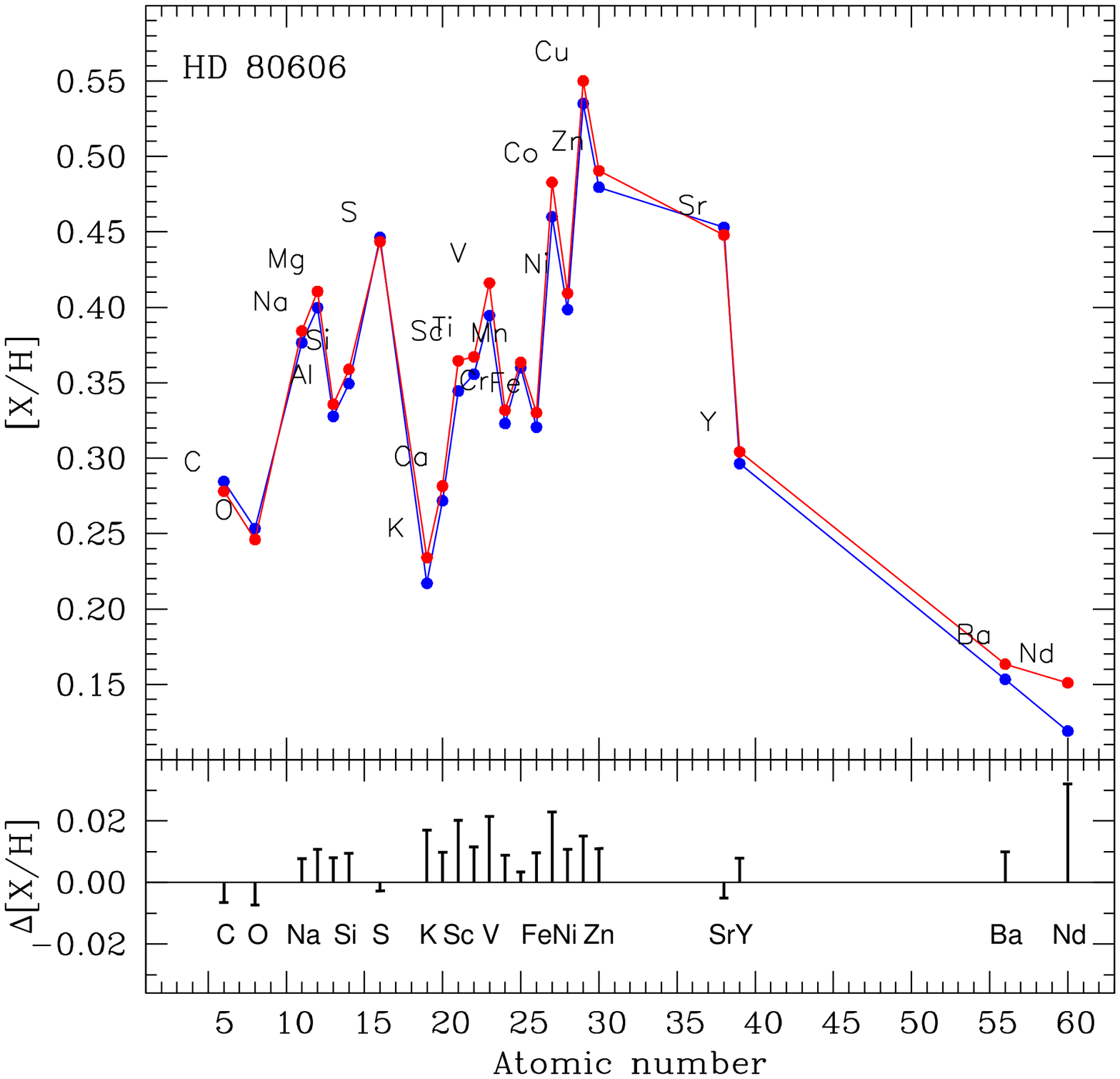}
\caption{Chemical pattern derived using the classical solar-scaled method (red) and the new method (blue).
The small lower panels show the abundance differences between both methods, being a 
chemical pattern difference. Upper and lower figures correspond to the stars HAT-P-4 and HD 80606. }
\label{fig.patterns}
\end{figure}

In Figure \ref{fig.patterns} we show two examples of chemical pattern derived
using the classical solar-scaled method (red) and the doubly-iterated method (blue),
for the stars HAT-P-4 and HD 80606 (upper and lower figures).
For each star, the lower panel shows the abundance differences between the methods
(as new method $-$ solar-scaled).
We note that some of these differences are notably higher than the $\sim$0.01 dex of the [Fe/H]
(e.g. for La, Ce, Nd, Sm and Dy in HAT-P-4), reaching up to $\sim$0.03 dex. 
Other species shows even a contrary or negative difference (such as C and O in the same panel).
This behaviour of the different chemical elements is also seen in other stars
(see e.g. HD 80606, lower panel of Figure \ref{fig.patterns}).
In principle, we can consider these differences as a "chemical pattern difference" derived from
the use of one method or another (to our knowledge, showed for the first time in this work).
However, we caution that this pattern difference could change from star to star, depending
on their specific opacities (composition) and fundamental parameters.

We present in the Table \ref{tab.params} the final stellar parameters derived using the new scheme proposed here,
together with the individual errors showed as {T$_{\rm{eff}}\pm \xi_{Teff}$}, {log g $\pm \xi_{log g}$}, etc.
The errors were derived following the same procedure of previous works, taking into
account independent and covariance terms in the error propagation \citep[see e.g. Section 3 of ][ for more details]{saffe15}.
In addition, between parentheses we show the difference between both procedures (as new method $-$ solar-scaled method).
The higher differences in T$_{\rm{eff}}$, {log g}, [Fe/H] and v$_{\rm{micro}}$ amount to
(absolute values) 26 K, 0.05 dex, 0.020 dex and 0.05 km/s.
However, it is interesting to separate the results of main-sequence and giant stars, which seem to show
a slightly different behaviour.
For instance, there is no evolved star with a T$_{\rm{eff}}$ difference higher than $\sim$10 K, while many main-sequence
objects are above this value. A similar effect is seen in the average v$_{\rm{micro}}$ differences: there is no giant star
with a v$_{\rm{micro}}$ difference higher than 0.02 km/s, while many main-sequence objects are above this value.
On the other hand, there is no clear difference in the average difference values of main-sequence
and giant stars when comparing log g and [Fe/H] values.
We also note that the higher difference in T$_{\rm{eff}}$ corresponds to a main-sequence star (HAT-P-4 with 26 K), while the
higher difference in [Fe/H] corresponds to an evolved object (HD 32887 with 0.020 dex).

In general, the differences obtained between the solar-scaled and the new method, are comparable to the individual
dispersions of the values of the stellar parameters. 
In particular, we note that some [Fe/H] differences are higher or similar than the dispersions.
For the other three stellar parameters (T$_{\rm{eff}}$, log g and v$_{\rm{micro}}$) the differences are
usually lower than the individual dispersions. This means that the use of the non-solar-scaled method
is not neccesarily mandatory for every determination of stellar parameters. However, there are some
physical processes that claim a higher precision in order to be well described.
We discuss specifically the significance of these differences in the Section 5.1.

\begin{figure*}
\centering
\includegraphics[width=8cm]{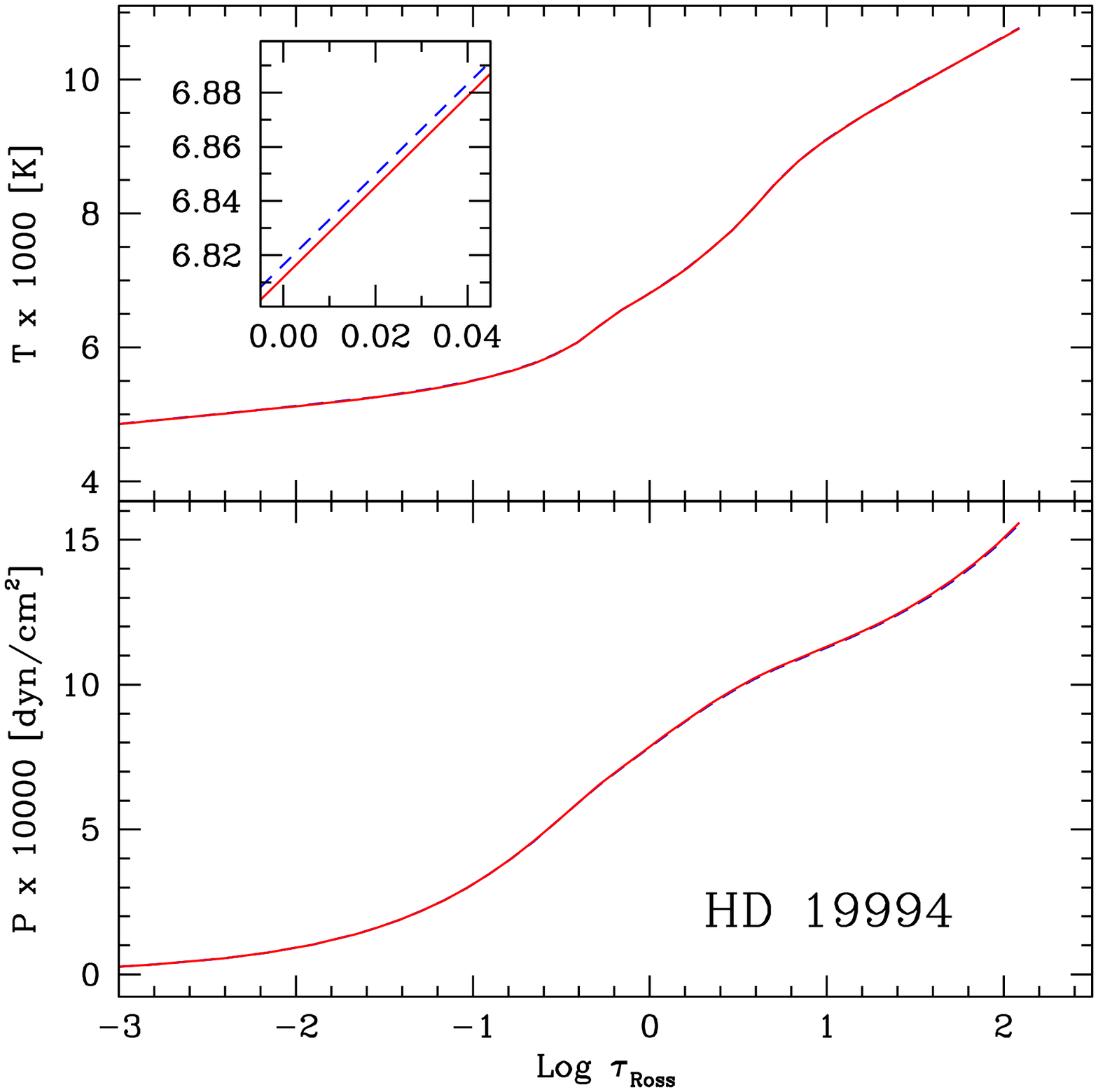}
\includegraphics[width=8cm]{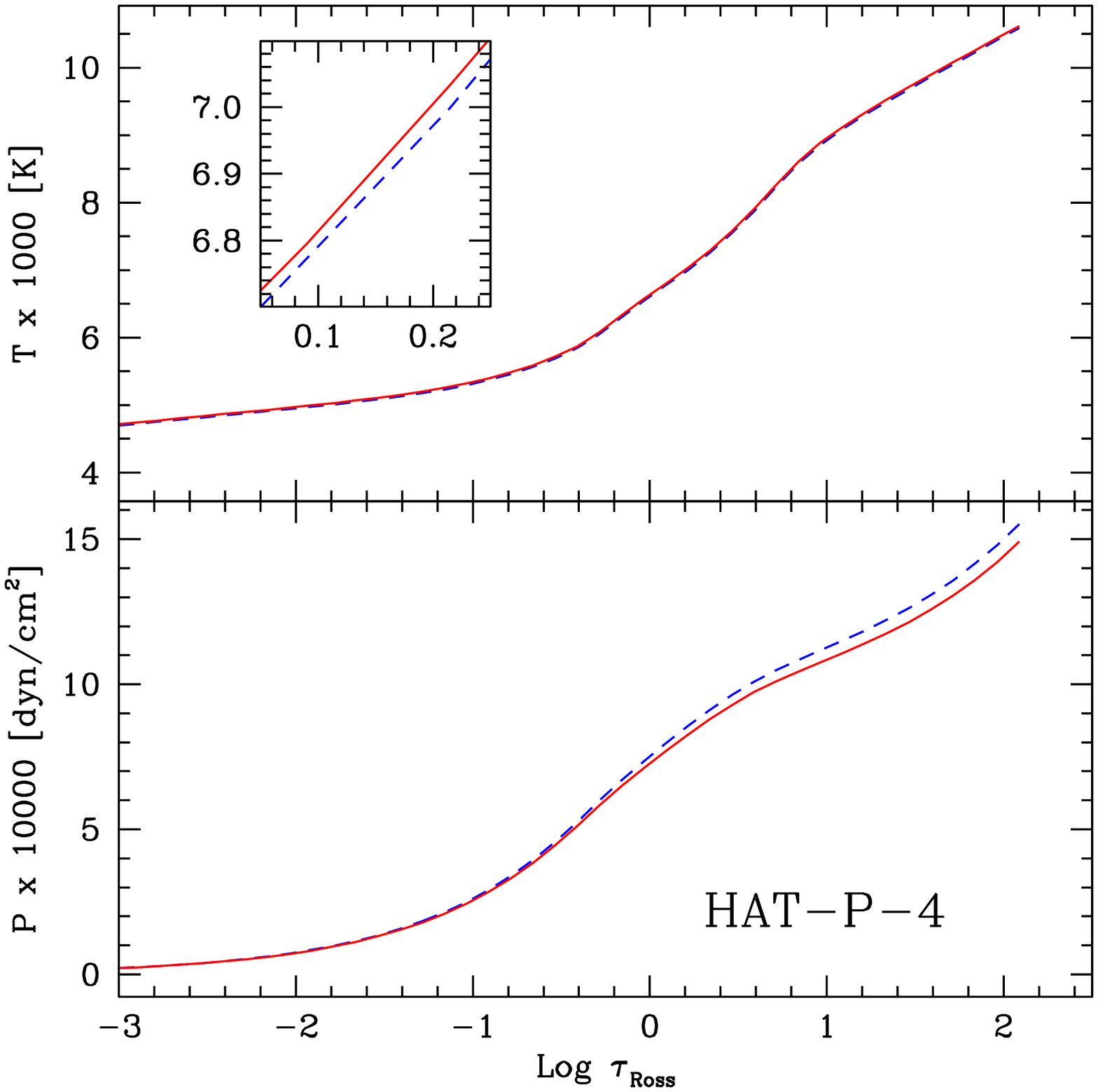}
\includegraphics[width=8cm]{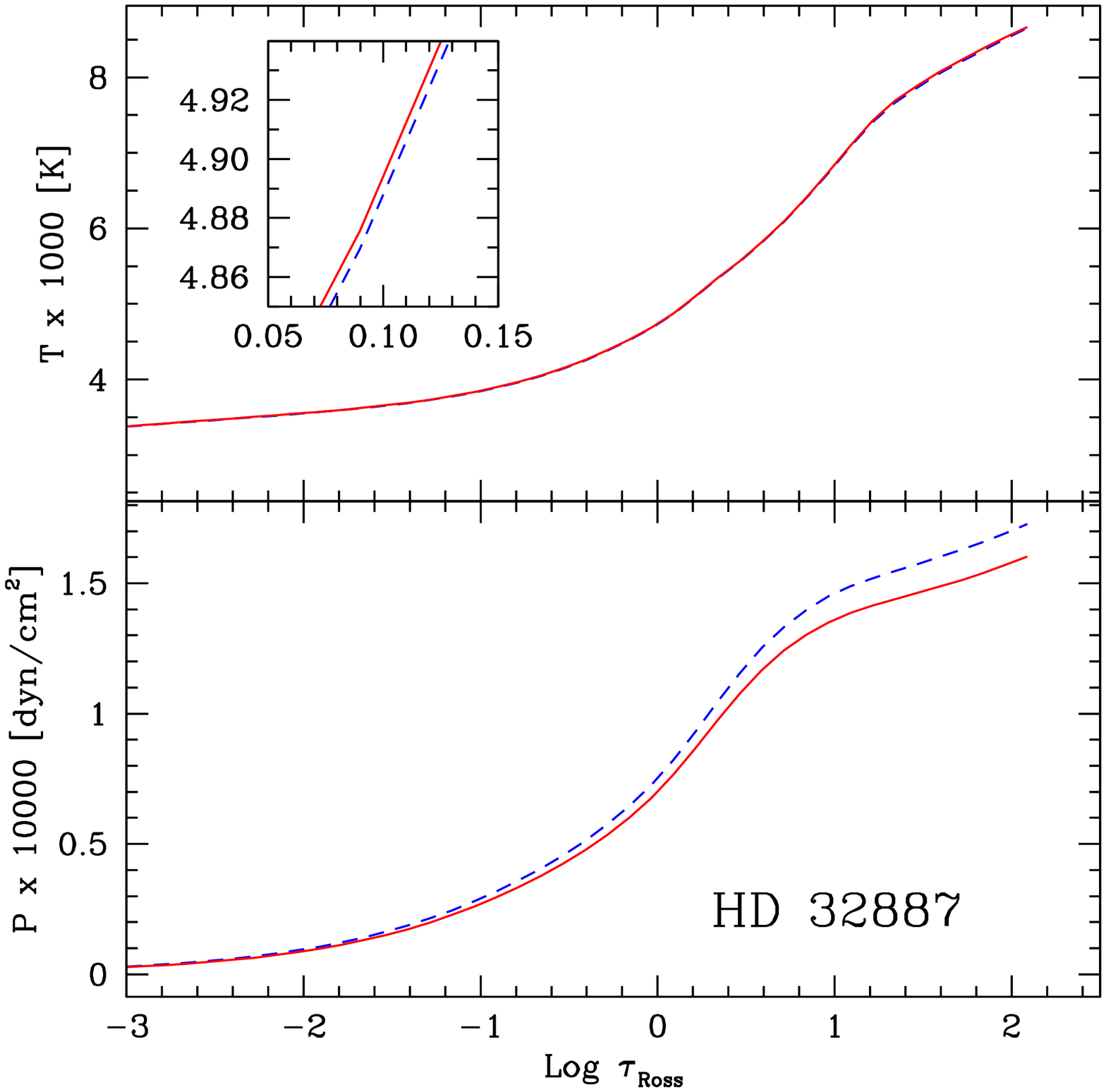}
\includegraphics[width=8cm]{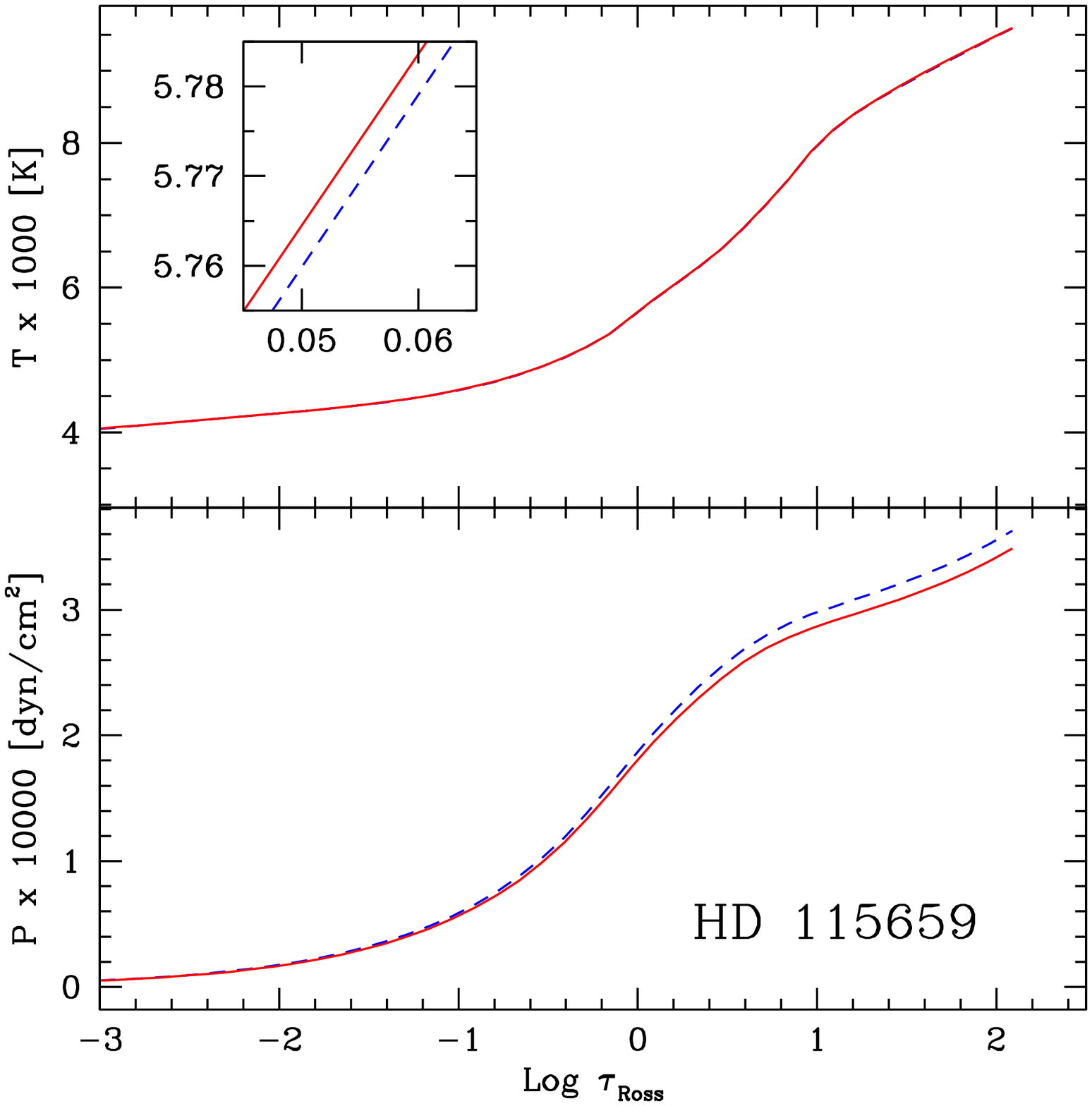}
\caption{Comparison of solar-scaled (blue dashed lines) vs non-solar-scaled (red continuous lines)
ATLAS12 model atmospheres. The example shows temperature and pressure distributions of four different
stars (see text for details).}
\label{fig.structure}
\end{figure*}

Given the differences found in stellar parameters between both methods showed in the Table \ref{tab.params},
we should expect small differences (although not identical) model structures derived using both methods.
We present in the Figure \ref{fig.structure} four examples comparing solar-scaled (blue dashed lines)
and non-solar-scaled (red continuous lines) ATLAS12 model atmospheres. This Figure include
two main-sequence stars (HD 19994, HAT-P-4) and two giant stars (HD 32887, HD 115659).
For each star, we show the temperature T and pressure P both as a function of the Rosseland
depth $\tau_{Ross}$, as derived from the ATLAS12 model atmospheres. 
We included intentionally the star HD 19994 in these plots, which shows one of the lowest differences in
stellar parameters when derived using both methods (see Table \ref{tab.params}).
The temperature distributions using both methods are similar in general. However, the small insets present a
zoom of the temperature distribution, showing indeed that the distributions are not identical: HD 19994 shows
a slightly higher T for the solar-scaled method than the new method, while the other three stars shows
a lower T for the solar-scaled method i.e. the contrary difference. This corresponds directly to the slightly higher
and lower T$_{\rm{eff}}$ respectively, obtained with both methods for these stars (see Table \ref{tab.params}).
In other words, a slightly higher T distribution corresponds to a slightly higher T$_{\rm{eff}}$.
In the same Figure \ref{fig.structure}, we also see that the pressure distributions show a more noticeable
difference when calculated using both methods. In particular, the lowest difference corresponds in this example to
HD 19994, showing one of the lowest differences in parameters in the Table \ref{tab.params}.
We note that the small differences found between the models showed in the Figure \ref{fig.structure}
include the range -1 $<$ log $\tau_{Ross}$ < 1 where we expect that most non-saturated spectral
lines should form. 
Then, the plots of Figure \ref{fig.structure} show that the structure of the model
atmospheres, as expected, are similar but not identical (even for stars such as HD 19994), and the differences
correspond to the differences found in stellar parameters showed in the Table \ref{tab.params}.

\begin{figure}
\centering
\includegraphics[width=8cm]{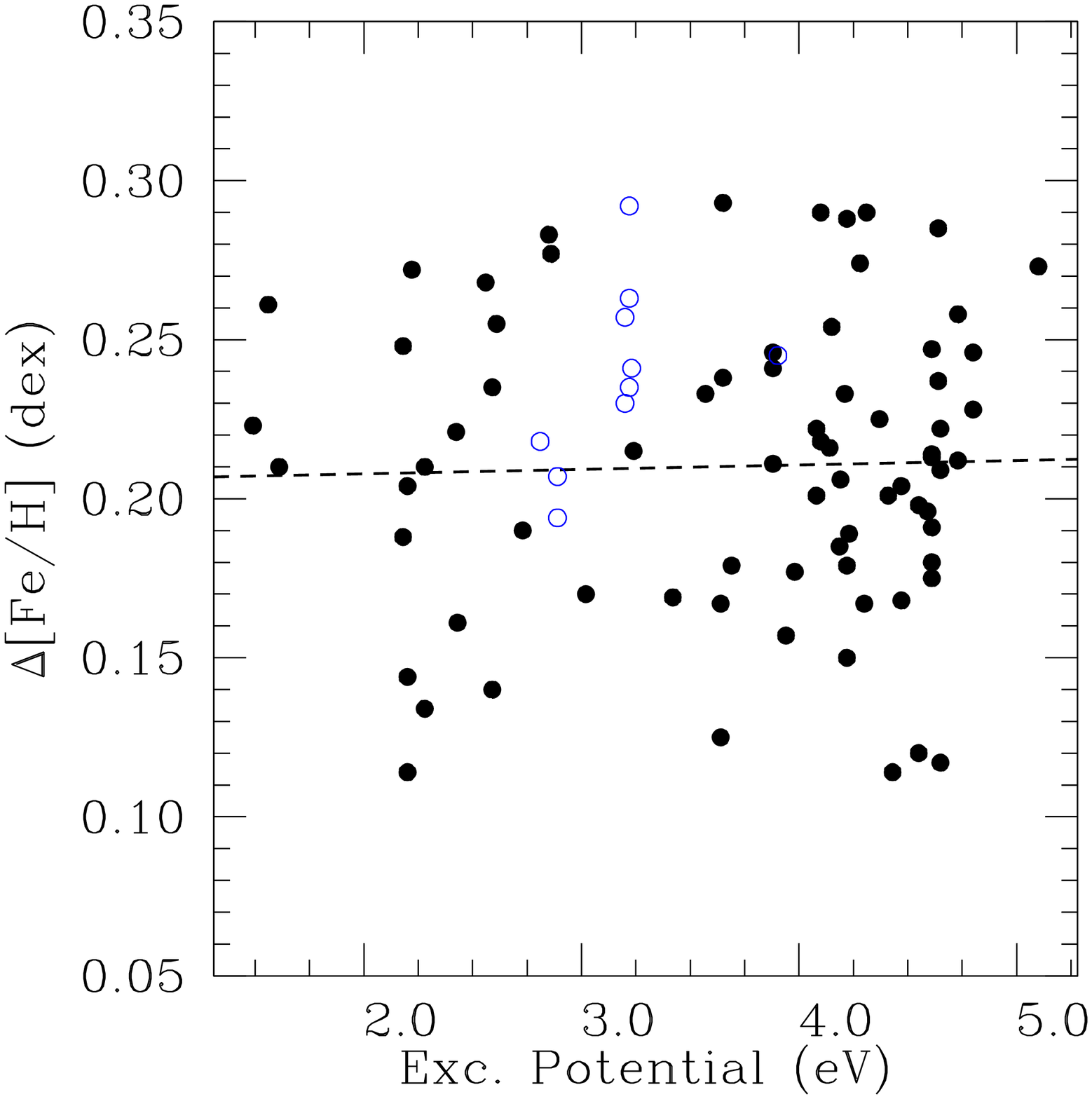}
\includegraphics[width=8cm]{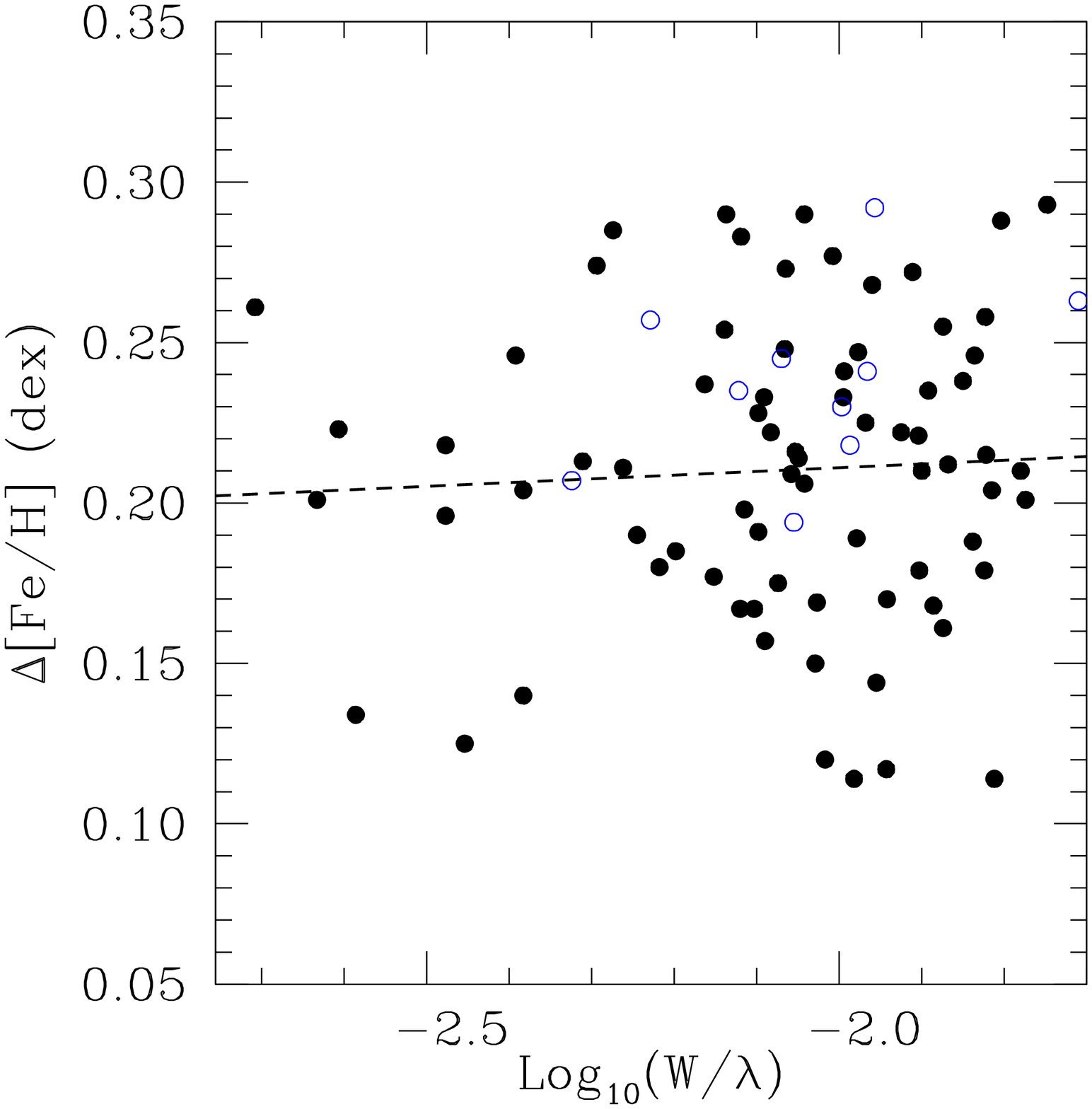}
\caption{Iron abundance vs excitation potential (upper panel) 
and iron abundance vs reduced EW (lower panel) for the star HAT-P-4,
derived with non-solar-scaled opacities but applying the solution of the
solar-scaled method. Filled and empty points correspond to Fe I and Fe II,
respectively. The dashed line is a linear fit to the abundance values.
See text for more details.}
\label{fig.non-equil}%
\end{figure}

We also performed the following experiment.
Using non-solar-scaled opacities, we derived the abundances of Fe I and
Fe II but applying the solution found with the solar-scaled method.
We present in the Figure \ref{fig.non-equil} the corresponding results,
showing the iron abundance vs excitation potential and iron abundance vs
reduced EW for the star HAT-P-4 (upper and lower panels).
Filled and empty points correspond to Fe I and Fe II, while the dashed
line shows a linear fit to the abundance values.
Inspecting the plots, it is clear that the four conditions required to
find an appropriate solution for the model atmosphere (see Section 4) are
not satisfied simultaneously. The average Fe II abundance is higher than the Fe I average
(0.24 and 0.21 dex, respectively), with 8/10 Fe II lines above the Fe I average.
Also, the abundances do present a slight trend with the reduced equivalent
widths (lower panel). Even admitting a null slope in both panels of
Figure \ref{fig.non-equil}, the fact that only one condition is not verified
(e.g. the ionization balance), is enough to show that the stellar parameters 
using the solar-scaled method indeed were not exactly derived, and then a
new solution could improve the previous one. And this new solution will need to
change the four stellar parameters, not only one, because the four mentioned conditions
are not independent between them. This task is done by the
downhill simplex method within the FUNDPAR program, using non-solar-scaled
opacities as explained in the Section 4.

From Table \ref{tab.params}, we note that four main-sequence stars show a negative difference
in T$_{\rm{eff}}$ between both methods (HD 19994, HD 221287, HD 96568 and HD 128898),
being the stars with the highest T$_{\rm{eff}}$ in the {{main-sequence}} sample.
Then, we wonder if the differences between both methods depend only on one parameter such as e.g. T$_{\rm{eff}}$.
HAT-P-4 and HD 80606 differ both by $\sim$+25 K when using both methods but having a T$_{\rm{eff}}$
difference of $\sim$485 K. Also, TYC 2567-744-1 and HD 80607 differ by $\sim$+15 K between
both methods and have a T$_{\rm{eff}}$ difference of $\sim$550 K. Then, we cannot adopt solely T$_{\rm{eff}}$
as a proxy for the differences between both methods, although it seems to play a role.
Due to the explicit calculation of the opacities for specific composition, we should expect that
the differences also depend on the detailed chemical composition of each target.

We note that the differences in [Fe/H] between both methods are very similar when considering two stars
in the same binary system. For instance, HD 80606 and HD 80607 present differences of +0.009 and +0.006 dex,
while $\zeta^{1}$ Ret and $\zeta^{2}$ Ret present differences of +0.010 and +0.009 dex. 
This implies that the mutual difference in [Fe/H] between binary components is approximately preserved
when considering two very similar stars.

The last two columns in the Table \ref{tab.params} show the slope of abundance vs condensation temperature 
T$_{c}$, using abundances of the classical solar-scaled method (slope s$_{\rm{classic}}$) and abundances
with the new scheme (slope s$_{\rm{new}}$). Condensation temperatures were taken from the 50\% T$_{c}$ values derived
by \citet{lodders03} for a solar system gas with [Fe/H]=0.
In particular for the binary systems of Table \ref{tab.params}, if one star presents a higher s$_{\rm{classic}}$ value
than its binary companion, then the same star also presents a higher slope s$_{\rm{new}}$ in the new method.
Then, the T$_{c}$ trends of previous works (SA15, SA16, SA17) are verified.
However, we consider that this is not neccesarily the rule, because both slopes s$_{\rm{classic}}$ and s$_{\rm{new}}$
present small but noticeable differences between them in general,
differences which are comparable to the error of the slopes and then cannot be totally ignored.

From Table \ref{tab.params}, the values of s$_{\rm{new}}$ and s$_{\rm{classic}}$ values are similar
within their errors, showing in general lower differences for giants than for main-sequence stars.
Some slopes of the T$_{c}$ trends remains almost identical (e.g. s$_{\rm{new}}$ and s$_{\rm{classic}}$ of HD 2114
are 3.03$\pm$1.87 and 3.06$\pm$1.87), while other T$_{c}$ trends could change up to $\sim$100\% of their
original slopes (e.g. s$_{\rm{new}}$ and s$_{\rm{classic}}$ of HD 80606 are 2.38$\pm$0.94 and 4.17$\pm$0.85). 
We note that main-sequence stars with lower T$_{\rm{eff}}$ seem to show
higher s$_{\rm{new}}$ values than s$_{\rm{classic}}$, while stars with T$_{\rm{eff}}$ higher than $\sim$6200 K
seem to show the contrary effect. For giant stars, those stars with T$_{\rm{eff}}$ lower than
$\sim$4500 K seem to show s$_{\rm{new}}$ greater than s$_{\rm{classic}}$, while for higher T$_{\rm{eff}}$ the
behaviour is less clear. Then, the exact difference between the T$_{c}$ trends depends on the fundamental
parameters of the stars and of their chemical patterns.

\begin{table*}
\centering
\caption{Stellar parameters derived using the new doubly-iterated method.
The difference between both procedures is indicated with parentheses (as new method $-$ solar-scaled method).}
\hskip -0.10in
\scriptsize
\begin{tabular}{lcccccc}
\hline
\hline
Star & T$_{\rm{eff}}$ & {log g} & [Fe/H] & v$_{\rm{micro}}$ & s$_{\rm{classic}}$ & s$_{\rm{new}}$ \\
     & [K]  & [dex]& [dex]  & [km/s] & [10$^{-5}$ dex/K] & [10$^{-5}$ dex/K] \\
\hline
Main-sequence stars\\
\hline
HAT-P-4         & 6064$\pm$35 ($+$26) & 4.34$\pm$0.08 ($-$0.01) & $+$0.220$\pm$0.005 ($+$0.009) & 1.30$\pm$0.09 ($+$0.04) & $+$12.37$\pm$1.53 & $+$13.70$\pm$1.50 \\
TYC 2567-744-1  & 6055$\pm$37 ($+$15) & 4.37$\pm$0.07 ($-$0.01) & $+$0.104$\pm$0.006 ($+$0.012) & 1.22$\pm$0.10 ($+$0.02) & $+$3.98$\pm$0.81  & $+$5.29$\pm$0.95 \\
HD 80606        & 5579$\pm$28 ($+$25) & 4.31$\pm$0.10 ($+$0.02) & $+$0.268$\pm$0.002 ($+$0.009) & 0.91$\pm$0.07 ($+$0.04) & $+$2.38$\pm$0.94  & $+$4.17$\pm$0.85 \\
HD 80607        & 5505$\pm$34 ($+$16) & 4.29$\pm$0.11 ($+$0.01) & $+$0.249$\pm$0.004 ($+$0.006) & 0.88$\pm$0.09 ($+$0.03) & $+$3.23$\pm$0.64  & $+$4.30$\pm$0.55 \\
$\zeta^{1}$ Ret & 5725$\pm$29 ($+$11) & 4.50$\pm$0.05 ($+$0.00) & $-$0.251$\pm$0.003 ($+$0.010) & 0.94$\pm$0.05 ($+$0.05) & $+$2.67$\pm$1.26  & $+$2.81$\pm$1.21 \\
$\zeta^{2}$ Ret & 5874$\pm$27 ($+$13) & 4.52$\pm$0.07 ($-$0.01) & $-$0.270$\pm$0.004 ($+$0.009) & 1.08$\pm$0.06 ($+$0.05) & $-$2.31$\pm$1.37  & $-$0.84$\pm$1.34 \\
HD 19994        & 6245$\pm$39 ($-$4)  & 4.42$\pm$0.15 ($+$0.00) & $+$0.228$\pm$0.010 ($-$0.001) & 1.38$\pm$0.11 ($-$0.01) & $-$2.82$\pm$1.37  & $-$3.51$\pm$1.37 \\ 
HD 221287       & 6340$\pm$40 ($-$6)  & 4.63$\pm$0.13 ($+$0.01) & $+$0.007$\pm$0.010 ($-$0.001) & 1.28$\pm$0.12 ($-$0.01) & $+$7.60$\pm$1.48  & $+$7.47$\pm$1.47 \\
HD 96568        & 8333$\pm$39 ($-$14) & 3.59$\pm$0.09 ($+$0.04) & $-$0.081$\pm$0.005 ($+$0.010) & 1.54$\pm$0.08 ($+$0.02) & $+$6.56$\pm$4.77  & $+$6.40$\pm$4.77 \\
HD 128898       & 8035$\pm$27 ($-$8)  & 4.64$\pm$0.09 ($+$0.02) & $+$0.099$\pm$0.008 ($+$0.008) & 1.59$\pm$0.11 ($-$0.03) & $+$7.99$\pm$4.82  & $+$6.59$\pm$4.84 \\
\hline
Evolved stars\\
\hline
HD 2114         & 5308$\pm$35 ($-$8)  & 2.87$\pm$0.12 ($-$0.03) & $-$0.047$\pm$0.010 ($+$0.000) & 1.86$\pm$0.10 ($-$0.01) & $+$3.06$\pm$1.87  & $+$3.03$\pm$1.87 \\
HD 10761        & 5036$\pm$32 ($+$8)  & 2.74$\pm$0.10 ($-$0.03) & $+$0.018$\pm$0.007 ($+$0.005) & 1.49$\pm$0.09 ($+$0.01) & $+$7.04$\pm$2.50  & $+$7.22$\pm$2.46 \\
HD 28305        & 5020$\pm$41 ($+$1)  & 3.00$\pm$0.11 ($+$0.01) & $+$0.157$\pm$0.006 ($+$0.004) & 1.72$\pm$0.11 ($-$0.01) & $+$9.36$\pm$2.68  & $+$9.42$\pm$2.66 \\
HD 32887        & 4271$\pm$39 ($+$6)  & 1.91$\pm$0.12 ($-$0.04) & $-$0.169$\pm$0.006 ($+$0.020) & 1.57$\pm$0.08 ($+$0.00) & $+$6.62$\pm$3.10  & $+$7.47$\pm$3.10 \\
HD 43023        & 5069$\pm$29 ($-$7)  & 3.05$\pm$0.11 ($+$0.01) & $-$0.020$\pm$0.004 ($-$0.001) & 1.25$\pm$0.05 ($-$0.02) & $+$4.44$\pm$1.82  & $+$3.64$\pm$1.81 \\
HD 50778        & 4131$\pm$39 ($+$7)  & 1.62$\pm$0.16 ($-$0.05) & $-$0.431$\pm$0.007 ($+$0.014) & 1.60$\pm$0.10 ($+$0.00) & $+$4.42$\pm$3.08  & $+$5.34$\pm$3.10 \\
HD 85444        & 5179$\pm$31 ($+$2)  & 2.97$\pm$0.14 ($-$0.01) & $+$0.051$\pm$0.010 ($+$0.004) & 1.47$\pm$0.07 ($+$0.00) & $+$5.61$\pm$2.02  & $+$7.72$\pm$2.08 \\
HD 109379       & 5240$\pm$36 ($+$5)  & 2.75$\pm$0.15 ($+$0.00) & $-$0.003$\pm$0.006 ($+$0.005) & 1.75$\pm$0.13 ($+$0.00) & $-$1.80$\pm$2.35  & $-$1.50$\pm$2.35 \\
HD 115659       & 5159$\pm$29 ($+$4)  & 2.94$\pm$0.13 ($-$0.02) & $+$0.066$\pm$0.008 ($+$0.004) & 1.49$\pm$0.07 ($+$0.00) & $+$8.74$\pm$2.28  & $+$9.03$\pm$2.26 \\
HD 152334       & 4291$\pm$37 ($+$5)  & 2.11$\pm$0.12 ($-$0.01) & $-$0.072$\pm$0.005 ($+$0.010) & 1.41$\pm$0.11 ($-$0.01) & $+$5.92$\pm$3.04  & $+$6.67$\pm$3.56 \\
\hline
\end{tabular}
\normalsize
\label{tab.params}
\end{table*}

\subsection{Solar-scaled or not: are the differences significant?}

At first view, a difference of 0.01 dex in metallicity or 15 K in T$_{\rm{eff}}$ does not seem to be significant. 
For instance, a number of works show that giant planets form preferentially around metal-rich stars
\citep[e.g. ][]{santos04,santos05,fischer-valenti05}, which is called the giant planet-metallicity correlation.
In these statistical works with hundreths of objects, an individual dispersion in [Fe/H] of
0.01 or 0.02 dex would not be significant.
However, exoplanet host stars are a fossil record of planet formation, beyond the
giant planet-metallicity correlation.
Using a high-precision abundance analysis, \citet{melendez09} found a lack of refractories
in the atmosphere of the Sun, when compared to the average abundances of 11 solar twins. 
The authors found a trend between the abundances and T$_{c}$ of the different
chemical species, as a possible signature of rocky planet formation. They proposed that the refractory
elements absent in the Sun's atmosphere were used to create rocky planets and the nuclei of giant planets.
This idea was followed by a number of works in literature \citep[see e.g. ][]{ramirez10,tucci-maia14,saffe15,saffe16,saffe17}.
The detection of this planetary chemical signature is a very challenging task,
requiring a careful and detailed analysis of the data, as we explain below.

The complete T$_{c}$ trend detected by \citet{tucci-maia14} for the stars of the binary system 16 Cyg,
covers a range of only 0.04 dex between the maximum and minimum abundance values of 19 different chemical species
(see their Figure 3). 
We showed previously (Figure \ref{fig.patterns}) that a number of species suffer an abundance difference
(as new method $-$ solar-scaled) higher than the $\sim$0.01 dex of the [Fe/H], reaching up to
$\sim$0.03 dex for e.g. Nd or Sm in HAT-P-4, with some elements showing even the contrary difference
(such as C and O).
Then, it is clear that the inclusion or not of this effect will have an important impact in the detection of a possible
T$_{c}$ trend, in objects such as the stars of the 16 Cyg binary system, where the complete trend covers only 0.04 dex.
This shows that the detection of a possible T$_{c}$ trend (as a chemical signature of planet formation)
requires the maximum achievable precision in both stellar parameters and abundances.
In general, we can say that those high-precision studies whose results depend on the mutual comparison of
different chemical species should prefer the non-solar-scaled method.

It is also notable that our Sun is considered a typical star in the context of the giant
planet-metallicity correlation, however the same Sun do shows a clear T$_{c}$ trend when compared to
solar twins \citep{melendez09,ramirez10} using very high-precision abundances.
This apparent dichotomy derive in part from the precision reached in these different works.
It would be very difficult (if not impossible) to detect a slight T$_{c}$ trend by applying
only standard techniques (e.g. non-differential) rather than high-precision methods of stellar parameters
derivation.

The search of a possible chemical signature of planet formation is not the only reason to 
pursuit high-precison stellar parameters. Several examples in the literature shows that, for instance,
a precise planetary characterization depends on the precise derivation of stellar parameters.
For instance, \citet{bedell17} revised the stellar parameters of the host star
Kepler-11 using a high-precision spectroscopic method and showed that the planet
densities raised between 20\% and 95\% per planet compared to previous works.
Other example corresponds to the derivation of the planetary radius which
depends on the stellar radius, which in turn is derived from the fundamental
parameters T$_{\rm{eff}}$, log g, [Fe/H] and v$_{\rm{micro}}$ \citep[see e.g. ][]{johnson17}.
Notably, the improved derivation of parameters allowed the detection of a gap in
the radius distribution of small planets \citep[R $<$ 2 R$_{Earth}$, ][]{fulton17},
showing the possible presence of two different populations of planets
(rocky planets and Neptune-like planets). Then, the importance of precise stellar parameters
is evident. Also, we can consider the missing mass of refractory material in the atmosphere of
the stars due to the planet formation process, estimated using the models of \citet{chambers10}.
Using the solar-scaled results for the star HAT-P-4 we obtain M$_{rock}$ $=$ 7.2 $\pm$ 0.4 M$_{Earth}$,
while using the non-solar-scaled values we derive M$_{rock}$ $=$ 8.6 $\pm$ 0.4 M$_{Earth}$ i.e. a 20\%
of difference. This error could be avoided using the new estimation of parameters.
Then, it is clear the need for reach the highest possible precision in these works.

Following the previous considerations, we can summarize as follows.
If we are working with a large sample of stars, trying only to detect e.g. a global difference in metallicity
of $\sim$0.20 dex between two subgroups, then an individual dispersion of $\sim$0.01 dex does not
seem to be significant. On the other hand, if we are trying to detect e.g. a relative difference in
abundances of different species (such as a T$_{c}$ trend) or e.g. the missing mass of refractory material
due to the planet formation process, then it is worthwhile to use the highest possible precision.
It is also important to keep in mind that some chemical species show differences higher than that of [Fe/H].
We caution that the use of a lower precision will certainly tend to mask or smooth out the signal of
any physical process which requires a higher precision to be detected.

\subsection{A bias in studies of many stars?}

From Table \ref{tab.params}, although T$_{\rm{eff}}$ seems to play a role, it is not clear that
a single parameter determines the differences between the two methods. However, 
we can suppose that a group of stars in the solar neighborhood presents
a chemical pattern approximately similar to the solar one, while other stars
have in general a non-solar-scaled pattern.
Suppose that we obtain stellar parameters for all stars using the classical two-step
method and the non-solar-scaled method. 
Then, those stars in the first group will show similar parameters derived with
both methods, given their chemical pattern approximately solar.
However, those stars in the second group will present a difference in their
parameters, which gradually increases as stars present patterns farther and farther
from the solar one. This would correspond to a possible small bias in the parameters,
depending on the chemical composition of the stars considered.
In addition, \citet{castelli05a} showed that the internal temperature distribution 
T vs log $\tau_{Rosseland}$ of ATLAS9 and ATLAS12 model atmospheres do differ with increasing T$_{\rm{eff}}$,
adopting in principle the same abundance pattern in both cases.
Then, we consider that a possible small bias cannot be totally ruled out.
We will address this important question in detail quantitatively in a next work.

\section{Conclusion}

We used non-solar-scaled opacities for a simultaneous derivation of
stellar parameters and chemical abundances in a sample of solar-type main-sequence and evolved
stars. To date, this is probably one of the more precise spectroscopic methods of stellar parameters
determination.
The difference in stellar parameters could amount to +26 K in T$_{\rm{eff}}$, 0.05 dex in {log g} and
0.020 dex in [Fe/H], when using solar-scaled vs non-solar-scaled methods.
We note that some chemical species could also show a variation higher than those of the [Fe/H],
and varying from one specie to another, obtaining a chemical pattern difference between
both methods. This means that T$_{c}$ trends could also present a variation.
The differences were derived using the full line-by-line differential technique
i.e. both for the derivation of stellar parameters and chemical abundances.
We consider that the use of non-solar-scaled opacities is not neccesarily mandatory e.g.
in statistical studies with large sample of stars. 
On the other hand, those high precision studies whose results depend on the mutual comparison of
different chemical species (such as a T$_{c}$ trend) should prefer the non-solar-scaled method.
In these cases, when modeling the atmosphere of the stars, the four stellar parameters
usually taken as (T$_{\rm{eff}}$, {log g}, [Fe/H], v$_{\rm{micro}}$) should in fact be considered
as (T$_{\rm{eff}}$, {log g}, chemical pattern, v$_{\rm{micro}}$), which is conceptually closer to
the real case.

\begin{acknowledgements}
M. F., F. M. L. and M.J.-A. acknowledge the financial support from CONICET in the forms of Post-Doctoral Fellowships.
We also thank the referee for their comments that greatly improved the paper.
\end{acknowledgements}

\end{document}